\documentclass[a4paper,11pt]{article}
\usepackage{latexsym}
\usepackage{amsmath, amsthm, amssymb, bbm}
\usepackage[mathscr]{eucal}
\usepackage{hyperref}
\usepackage{slashed}
\usepackage{graphicx}
\usepackage{authblk}

\theoremstyle{plain}

\theoremstyle{definition}

\theoremstyle{remark}

\newcommand{\rhs}{r.h.s.\ }

\newcommand{\wrt}{w.r.t.\ }
\newcommand{\cf}{cf.\ }

\newcommand{\ud}{\mathrm{d}}
\newcommand{\del}{\partial}

\newcommand{\betrag}[1]{{\lvert #1 \rvert}}

\newcommand{\Z}{\mathbb{Z}}

\newcommand{\order}{\mathcal{O}}

\newcommand{\eps}{\varepsilon}

\DeclareMathOperator{\tr}{tr}

\newcommand{\N}{\mathbb{N}}

\newcommand{\s}[1]{{\underline{#1}}}

\newcommand{\ret}{{\mathrm{ret}}}
\newcommand{\adv}{{\mathrm{adv}}}
\newcommand{\nabslash}{{\slashed \nabla}}

\begin{document}

\title{The current density in quantum electrodynamics in external potentials}
\author[1]{Jan Schlemmer\thanks{jan.schlemmer@univie.ac.at}}
\author[2]{Jochen Zahn\thanks{jochen.zahn@itp.uni-leipzig.de}}

\renewcommand\Affilfont{\itshape\small}
\affil[1]{Fakult\"at f\"ur Physik, Universit\"at Wien, Boltzmanngasse 5, 1090 Wien, Austria}
\affil[2]{Institut f\"ur Theoretische Physik, Universit\"at Leipzig, Br\"uderstr.~16, 04103~Leipzig, Germany}

\date{\today}

\maketitle

\begin{abstract}
We review different definitions of the current density for quantized fermions in the presence of an external electromagnetic field. Several deficiencies in the popular prescription due to Schwinger and the mode sum formula for static external potentials are pointed out. We argue that Dirac's method, which is the analog of the Hadamard point-splitting employed in quantum field theory in curved space-times, is conceptually the most satisfactory. As a concrete example, we discuss vacuum polarization and the stress-energy tensor for massless fermions in 1+1 dimension. Also a general formula for the vacuum polarization in static external potentials in 3+1 dimensions is derived.
\end{abstract}


\section{Introduction}

Quantum field theory in external potentials has received increasing interest in recent years, triggered in particular by the envisioned high intensity laser sources, \cf \cite{Dunne08} for a review, and the new facilities for heavy ion collisions, in which one might probe the chiral magnetic effect \cite{CME}. Further applications are vacuum polarization corrections to atomic energy levels, \cf the review \cite{MohrPlunienSoff98}. The topic has a long history, dating back to the seminal works of Dirac \cite{Dirac34} and Heisenberg and Euler \cite{EulerHeisenberg}. Schwinger basically brought the theory to its present form, by introducing the proper time formalism and the concept of the \emph{effective action} \cite{Schwinger51}. For textbooks and reviews on the subject, we refer to \cite{GreinerMullerRafelski, GribMamayevMostepanenko, FradkinGitmanShvartsman, DittrichReuter, Dunne04}.

There are two types of observables in quantum electrodynamics in external potential that one might be interested in, namely \emph{global} and \emph{local} ones. The prime example for a global observable is the expectation number of electron positron pairs produced by the external field. An example for a local observable would be vacuum polarization, i.e., the expectation value of the charge density in a static external potential. More generally, the local \emph{current density} $j^\mu(x)$ should be an observable.

In principle, both types of observables can be studied in the framework of effective actions. The particle production rate is related to the imaginary part of the effective action, whereas the expectation value of the current can be obtained by variation \wrt the external potential $A_\mu$. However, there is the practical difficulty that the effective action can be explictly computed only for very special forms of the external potential: To the best of our knowledge, the most general prescription to compute the effective action was given in \cite{DunneHall98}, where it is implicitly assumed that the external potential is analytic. But an analytic potential can not be varied locally, so the expectation value of the local current density can not be computed in that way.

On the conceptual side, it would be desirable to have a definition of the current density that is independent of the choice of a state. This would allow to evaluate it also in thermal states, or in situations where there is no preferred vacuum state, due to non-stationarity of the external fields in the asymptotic past (e.g., as for a plane wave).

Classically, the current density of a Dirac field is given by
\begin{equation}
\label{eq:current}
 j^\mu(x) = - e \bar \psi(x) \gamma^\mu \psi(x).
\end{equation}
Upon quantization, one meets the immediate problem that point-wise products of quantum fields are not well-defined. In the absence of external potentials, one typically proceeds via \emph{point-splitting} \wrt the vacuum two-point function, i.e., normal ordering. However, it is not clear how to replace the vacuum two-point function in the case of non-trivial external potentials.

The first proposal for a point-splitting definition of the current seems to be due to Dirac \cite{Dirac34}. He assumed that in a physically reasonable state $\omega$ the two-point function
\[
 \omega(\psi(x) \bar \psi(y))
\]
has the same singularities as the vacuum two-point function, but possibly with modified coefficients. The proposal for a renormalization prescription was then to subtract the singular part. As discussed below, this is, in modern terminology, the assumption that the two-point function is of \emph{Hadamard form} and that the correct renormalization prescription is by point-splitting \wrt the \emph{Hadamard parametrix}. This is the renormalization prescription that we are going to employ here. The main shortcomings of Dirac's treatment are the missing discussion of the inherent ambiguities and the only implicit determination of the coefficients of the singular part of the two-point function. Both were treated subsequently by Heisenberg \cite{Heisenberg34}, who showed that there is an ambiguity in the definition of the current, namely, one may add multiples of the external current (which corresponds to a charge renormalization). Dirac's method was also used by Heisenberg and Euler \cite{EulerHeisenberg}, Uehling \cite{Uehling35}, and Serber \cite{Serber35} in their seminal works on vacuum polarization. However, it seems to have been essentially abandoned afterwards, being superseded by Schwinger's prescription. The reason for this may have been practical: The singular part is explicitly given in position space, whereas the two-point function is typically expressed as an integral over modes, from which the smooth remainder is difficult to extract, at least beyond the approximation of first order in the external potential.\footnote{This is also evidenced by the fact that the actual calculations of Heisenberg and Euler \cite{EulerHeisenberg} proceed by a cut-off in momentum space, not by a point-splitting. They claim to also have done the calculation via point-splitting, but remark that it ``turned out to be very complicated.''} One of the results of the present paper is a method, applicable in the case of static external potentials, to subtract the singular part directly in the mode integral, thus allowing for more reliable numerical calculations.

Another definition of the local current was given by Schwinger \cite{Schwinger51}, who proposed to perform the limit of coinciding points in \eqref{eq:current} in a symmetric way, leading to cancellations of divergent parts. However, not all divergences are canceled in this way: As we will see, a logarithmic divergence remains, unless the external current vanishes. Furthermore, even if the external current vanishes, Schwinger's prescription in general depends on the way the limit of coinciding points is performed. It should also be noted that some authors seem to omit the parallel transport that is implicit in Schwinger's point-splitting procedure, \cf the footnote in \cite{Schwinger59}. Then, further divergences remain, and the result may fail to be gauge invariant. It should also be noted that, contrary to Dirac's method, Schwinger's prescription can not be used to renormalize the stress-energy tensor.

Finally, for the case of a static external potential, there is a definition for the vacuum polarization based on the mode expansion:
\begin{equation}
\label{eq:ModeSum}
 \rho(x) = \frac{e}{2} \left( \sum_+ \betrag{\psi_n(x)}^2 - \sum_- \betrag{\psi_n(x)}^2 \right).
\end{equation}
This expression seems to originate in the work of Wichmann and Kroll \cite{WichmannKroll56} and is widely used in textbooks and reviews, e.g., \cite{GreinerQED, MohrPlunienSoff98}. Here the first sum is over positive and the second sum over the negative frequency modes, with the $\psi_n$'s properly normalized. Unless the system is spatially confined, the sums have of course to be replaced by integrals. Furthermore, the individual sums are divergent. In typical situations, positive and negative frequency modes can be paired such that the remaining sum is convergent (this was done in \cite{AmbjornWolfram83} for the case of a charged scalar field). As we will show by an elementary example, this expression in general leads to incorrect results.
The same is true for an analogous expression for the total energy.

In the context of quantum field theory on curved spacetimes, one also encounters the problem of defining Wick powers, i.e., non-linear local obervables, in the absence of a preferred vacuum state. In recent years, there has been tremendous progress in that field, triggered by the application of the tools of \emph{microlocal analysis} \cite{Radzikowski} and by the introduction of the framework of \emph{locally covariant field theory} \cite{HollandsWaldWick, BrunettiFredenhagenVerch}. In the latter, one does not define a quantum field theory on one particular spacetime, but on all possible ones simultaneously, in a coherent way. The same applies to observables, such as the stress-energy tensor. This is non-trivial, as it necessitates a point-splitting, which is required to depend only on the local geometric data, in order to fulfill the condition of local covariance. The framework was applied to conceptual as well as practical problems, such as the proof of spin-statistics \cite{VerchSpinStatistics} and CPT \cite{HollandsPCT} theorems, the discussion of quantum energy inequalities \cite{FewsterPfenningQEILocCovI}, or to cosmological back-reaction effects \cite{DappiaggiFredenhagenPinamonti}. For a recent review, we refer to \cite{HollandsWaldReview}.

The framework of locally covariant field theory can be straightforwardly generalized to fields that are charged under a gauge group $G$, in the presence of an external potential \cite{LocCovDirac}. Specializing to Minkowski spacetime and $G = U(1)$, one obtains a framework for quantum field theory in external electromagnetic potentials, in which all observables are constructed in a locally gauge covariant manner. Considering the electromagnetic field as purely external, one is dealing with a free quantum theory, in which only Wick powers need to be renormalized. This proceeds via point-splitting \wrt the Hadamard parametrix, i.e., as in Dirac's prescription. We note that the well-known chiral anomalies can be understood (and also computed in a rather elementary way) as due to the Hadamard parametrix being a bi-solution to the equation of motion only up to smooth remainders \cite{ChiralFermions}. This may be seen as another indication of the appropriateness of the framework.

One aim of this paper is to review Dirac's point-splitting procedure in modern terms and to relate it to Schwinger's prescription and the mode sum formula \eqref{eq:ModeSum}. We also discuss the remaining ambiguities. This is done in the next section. In Section~\ref{sec:2d}, we apply the procedure in a case that is analytically tractable, a massless fermion in two space-time dimensions in a constant electric field and confined to a spatial interval. This example also nicely illustrates the failure of the mode sum formula \eqref{eq:ModeSum}. For applications in four space-time dimensions, one meets the inconvenience that the Hadamard parametrix is known in position space, whereas the two-point functions of interesting states are typically given by mode integrals, where it is difficult to subtract the position space singularity. In Section~\ref{sec:4d}, we propose a solution to this problem for the case of static external potentials, and exemplify it by considering the Coulomb potential. This method can also be modified for the case of a homogeneous time-dependent electric field, which is discussed in a separate publication \cite{SauterPotential}.

\subsection*{Notations and conventions}

We work with signature $(+, -, -, -)$. The electric charge is denoted by $e$ and the covariant derivative is given by $\nabla_\mu = \del_\mu + i e A_\mu$. Throughout, we work in rationalized natural units, i.e., $c=\hbar=1$ and $\del^\nu F_{\mu \nu} = J_\mu$. With the Lagrangian
\[
 L = \bar \psi (i \nabslash - m) \psi - \tfrac{1}{4} F_{\mu \nu} F^{\mu \nu},
\]
these conventions determine the sign in \eqref{eq:current}. When considering functions (or sections) in two space-time variables, the second one is often denoted by $x'$. Hence, a primed covariant derivative is implied to act on the second variable. The coinciding point limit of such a smooth function is denoted by square brackets, $[f](x) = f(x, x)$. The spatial part of a Lorentz vector $x$ is denoted by $\s{x}$.

\section{Hadamard two-point functions and the parametrix}

In the following, we assume, as Dirac \cite{Dirac34}, that the physically relevant states have two-point functions of \emph{Hadamard form}, i.e., they are, in four space-time dimensions, locally given by
\begin{multline}
\label{eq:HadamardForm0}
 \omega(\psi^B(x) \bar \psi_A(x')) = \underbrace{\frac{{W_1}_A^B(x,x')}{((x-x')^2_\eps)^2} + \frac{{W_2}_A^B(x,x')}{(x-x')^2_\eps} + {W_3}_A^B(x,x') \log \frac{-(x-x')^2_\eps}{\Lambda^2}}_{H_A^B(x,x')} \\ + {W_4}_A^B(x,x'),
\end{multline}
with $A$ and $B$ spinor and co-spinor indices,
\begin{equation}
\label{eq:i_epsilon}
 (x-x')^2_\eps = (x-x')^2 - i \eps (x-x')^0,
\end{equation}
$\Lambda$ a length scale, and smooth functions $W_i$. The sum of the first three terms, i.e., $H$, is usually called the \emph{Hadamard parametrix}. Before discussing it further, let us recall the motivation for the assumption that the two-point function is of the above form.

Dirac's motivation was that the vacuum two-point function is of this form, and that the introduction of an external potential should not change the structure of the divergences. Nowadays, one can give further motivation, in particular due to the lessons learned in QFT on curved space-times. To begin with, the Hadamard form propagates, i.e., if the two-point function is of Hadamard form in the neighborhood of some time-slice and the potential is smooth, then it is of Hadamard form everywhere.\footnote{The first such result was obtained in \cite{FSW78} for scalar fields and generalized to Dirac fields in \cite{SahlmannVerchHadamard}. A modern proof using the tools of microlocal analysis was given in \cite{SandersDirac}. It is straightforwardly generalizable to the case of non-trivial external potentials.} Hence, for a potential that is switched on at some finite time, the two-point function that coincides with the vacuum two-point function in the past is of Hadamard form. In particular, this can be used to show that a two-point function of Hadamard form always exists \cite{LocCovDirac}, by adapting the deformation argument of \cite{FNW81}. Another relevant result is that, on Minkowski space and for generic smooth static external potentials, the ground state is of Hadamard form \cite{WrochnaDirac}. Finally, in states with two-point function of Hadamard form and smooth truncated $n$-point functions, the higher moments of smeared Wick powers are finite \cite{BrunettiFredenhagenScalingDegree}.\footnote{A partial converse of this statement was recently proved \cite{FewsterVerchHadamard}.} In particular this means that the Hadamard condition ensures finite fluctuations of the smeared current density.

Let us now review the Hadamard form of the two-point function. As suggested by Radzikowski \cite{Radzikowski}, it can be characterized by the conditions\footnote{A further condition is that the two-point function is the two-point function of a state, i.e., positive, $\omega(\bar \psi(u) \psi(\bar u)) \geq 0$ for $u$ a test section of the spinor bundle.}
\begin{align}
\label{eq:Comm}
 \omega(\psi^B(x) \bar \psi_A(x')) + \omega(\bar \psi_A(x') \psi^B(x)) & = i S^B_{\ A}(x,x'), \\
\label{eq:Conj}
 \overline{\omega(\bar \psi(u) \psi(\bar v))} & = \omega(\bar \psi(v) \psi(\bar u)),
\end{align}
and the requirement that the two-point function is of positive (negative) frequency in its first (second) argument. The latter is made precise by a condition on the wave front set (for an introduction to this concept we refer to \cite{BrouderDangHelein}). In \eqref{eq:Comm}, $S = S^\adv - S^\ret$ is the difference of advanced and retarded propagator of the Dirac operator (including the external potential), and in \eqref{eq:Conj}, $u$ and $\bar v$ denote test sections of the spinor and the co-spinor bundle, with which $\bar \psi$ and $\psi$ are integrated. 

One can show \cite{SandersDirac, LocCovDirac}, that these requirements fix the two-point function up to a smooth part, denoted by $W_4$ in \eqref{eq:HadamardForm0}. Furthermore, the first three terms in \eqref{eq:HadamardForm0}, are constructed locally and covariantly out of the geometric data, or, more precisely, the coinciding point limits
\begin{equation}
\label{eq:InvariantTensors}
 [\nabla_{\mu_1} \dots \nabla_{\mu_k} \nabla'_{\nu_1} \dots \nabla'_{\nu_l} W_i],
\end{equation}
of covariant derivatives of the $W_i$, $1 \leq i \leq 3$, are tensors constructed out of the external field strength $F$.

The proposal of Dirac \cite{Dirac34}, which was later ``rediscovered'' in the context of QFT on curved spacetimes \cite{DeWitt75, Christensen76, AdlerLiebermanNg76, Wald77, HollandsWaldWick},\footnote{Marecki, apparently unaware of Dirac's work, but  inspired by QFT on curved space-time, ``reintroduced'' this into QED in external potentials \cite{MareckiThesis}.} is then to define non-linear local observables (Wick powers) by a point-splitting in which the Hadamard parametrix $H$, i.e., the first three terms in \eqref{eq:HadamardForm0}, are subtracted. Concretely, for the evaluation of the Wick power $\psi^B \bar \psi_A(x)$ in the state $\omega$, one obtains
\[
 \omega( \psi^B \bar \psi_A(x) ) = [{W_4}_A^B](x).
\]
In particular, this provides a definition of the current $j^\mu$, by tracing the above with $\gamma^\mu$. It is clear that for the computation of the coinciding point limit of $W_4$ only a finite number of the tensors \eqref{eq:InvariantTensors} have to be known (up to four derivatives for $W_1$, up to two derivatives for $W_2$ and no derivative for $W_3$). If the Wick power contains derivatives (as needed for the stress-energy tensor), more of these tensors have to be known, but always just a finite number for any fixed order of derivatives. 

Before we finally compute the tensors \eqref{eq:InvariantTensors} that are relevant for the definition of the current, let us briefly comment on the ambiguities in the definition of Wick powers. These were not treated by Dirac, but a detailed discussion in the context of scalar fields on curved space-time can be found in \cite{HollandsWaldWick, HollandsWaldStress}. One obvious ambiguity is the choice of the length scale $\Lambda$ in the logarithmic term in the parametrix (a change would amount to adding a multiple of $[{W_3}_A^B](x)$ to the Wick power $\psi^B \bar \psi_A(x)$). More generally, one may allow for the freedom to add to a Wick square any local tensor of the correct dimension.\footnote{This can be seen as adding to the Hadamard parametrix $H$ a smooth term that is locally and covariantly constructed out of the geometric data. 
} For the Wick square $\varphi^2$ in four space-time dimensions, these are multiples of $m^2$ and $R$. For the Wick square $\psi^B \bar \psi_A$ on Minkowski space, examples of such tensors would be $m (\gamma^\mu \gamma^\nu)_A^B F_{\mu \nu}$ or ${\gamma^\mu}_A^B \nabla^\nu F_{\mu \nu}$. This freedom, however, is constrained by the requirement that the current should be conserved. Once the Wick square is chosen such that the current $j_\mu$ is conserved, the only remaining ambiguity consists of adding multiples of the external current $J_\mu = \del^\nu F_{\mu \nu}$ \cite{LocCovDirac}. This corresponds to a charge renormalization. Furthermore, as we will see below, in four space-time dimensions, this can be implemented by a change of the length scale $\Lambda$.

\subsection{The explicit form of the parametrix}
\label{sec:ExplicitForm}

We want to explicitly compute the tensors \eqref{eq:InvariantTensors} that are relevant for the point-splitting renormalization of the current density. Readers not interested in the details of this calculation, may skip towards the end of the subsection.

Due to the condition \eqref{eq:Comm}, it is advantageous to consider the singular behavior of the retarded and advanced propagators $S^{\ret/\adv}$ of the Dirac operator $i \nabslash - m$. To obtain these, we follow the strategy of Dimock \cite{DimockDirac}: The operator
\[
 P = (i \nabslash - m) (- i \nabslash - m) = \nabla_\mu \nabla^\mu + \tfrac{i}{2} e \gamma^\mu \gamma^\nu F_{\mu \nu} + m^2
\]
is \emph{normally hyperbolic}, i.e., its principal symbol is the metric and there are no first order derivatives. For such operators, there are unique retarded and advanced propagators, which are formally given by \cite{BGP07}
\begin{equation}
\label{eq:DeltaRetFormal}
 \Delta^{\ret / \adv}(x,x') = \sum_{k=0}^\infty V_k(x,x') R^{\ret / \adv}_{2k+2}(x,x'),
\end{equation}
where the \emph{Hadamard coefficients} $V_k$ are smooth and the \emph{Riesz distributions} $R_j$ in four space-time dimensions are given by
\begin{align*}
 R^{\ret / \adv}_2(x,x') & = \tfrac{1}{2 \pi} \theta(\pm (x-x')^0) \delta((x-x')^2), \\
 R^{\ret / \adv}_{2 + 2 k}(x,x') & = \tfrac{1}{2^{2k+1} \pi k! (k-1)!} (x-x')^{2 (k-1)} \theta(\pm (x-x')^0) \theta((x-x')^2),
\end{align*}
where in the second line $k \geq 1$. The Hadamard coefficients can be determined recursively by solving the transport equation
\begin{equation}
\label{eq:TransportEquation}
 (x-x')^\mu \nabla_\mu V_k(x,x') + k V_k(x,x') = - k P V_{k-1}(x,x'),
\end{equation}
with the initial condition $V_0(x, x) = 1$. This proceeds by integration of $P V_{k-1}$ along the straight path from $x$ to $x'$. For example, one straightforwardly obtains that
\[
 V_0(x, x') = \exp \left(-i e \int_0^1 A_\mu(x' + t (x-x')) (x-x')^\mu \ud t \right)
\]
is the parallel transport along the straight path from $x'$ to $x$. For the determination of the tensors $W_i$, $1 \leq i \leq 3$, we will need the coinciding point limit of up to three covariant derivatives of $V_0$ and up to one covariant derivative of $V_1$. For the parallel transport, one easily obtains
\begin{align}
\label{eq:V0_0}
 [V_0] & = 1, \\
\label{eq:V0_1}
 [\nabla_\mu V_0] & = 0, \\
\label{eq:V0_2}
 [\nabla_\mu \nabla_\nu V_0] & = \tfrac{i}{2} e F_{\mu \nu}, \\
\label{eq:V0_3}
 [\nabla_\mu \nabla_\nu \nabla_\lambda V_0] & = \tfrac{i}{3} e \left( \del_\mu F_{\nu \lambda} + \del_\nu F_{\mu \lambda} \right).
\end{align}
For $V_1$, we obtain, using \eqref{eq:TransportEquation} and \eqref{eq:V0_0} -- \eqref{eq:V0_3}, 
\begin{align}
\label{eq:V1_0}
 [V_1] & = - [P V_0] = - \tfrac{i}{2} e \gamma^\mu \gamma^\nu F_{\mu \nu} - m^2, \\
\label{eq:V1_1}
 [\nabla_\mu V_1] & = - \tfrac{1}{2} [\nabla_\mu P V_0] = - \tfrac{i}{4} e \gamma^\lambda \gamma^\rho \del_\mu F_{\lambda \rho} - \tfrac{i}{6} e \del^\lambda F_{\mu \lambda}.
\end{align}
It is clear that in principle one can compute coinciding point limits of arbitrarily many covariant derivatives of $V_k$ in this way.

The expansion \eqref{eq:DeltaRetFormal} is formal in the sense that the series does in general not converge; however, it is asymptotic in the sense that if it is truncated after $k=N$, then the difference $\Delta^{\ret / \adv} - \Delta_N^{\ret / \adv}$ of the actual and the truncated propagator is of regularity $C^{N-1}$ and can be bounded by $\betrag{ (x-x')^{2 (N-1)}}$ \cite{BGP07}. As we are ultimately interested in the limit of coinciding points, the formal expansion \eqref{eq:DeltaRetFormal} is completely sufficient for our purposes, also for the similarly defined parametrix.

The retarded/advanced propagator for the Dirac operator $i \nabslash - m$ is now defined as
\begin{equation}
\label{eq:S_ret}
 S^{\ret / \adv} = (- i \nabslash - m) \Delta^{\ret / \adv}.
\end{equation}
It is obvious that these have the required support property and that
\[
 (i \nabslash - m ) S^{\ret / \adv} = \delta.
\]
That also
\[
 S^{\ret / \adv} \circ (i \nabslash - m ) = \delta
\]
was shown by Dimock \cite{DimockDirac}. In particular, it follows that
\begin{equation}
\label{eq:S_symmetry}
 S^{\ret / \adv}(x,x') = (- i \nabslash - m) \Delta^{\ret / \adv}(x,x') = ( i {\nabslash^*}' - m) \Delta^{\ret / \adv}(x,x'),
\end{equation}
where the operator ${\nabslash^*}'$ is the conjugate Dirac operator (for charge $-e$) acting from the right on the second variable (the change of sign is due to partial integration), i.e.,
\[
 {\nabslash^*}' F(x,x') = (\del'_\mu - i e A_\mu(x')) F(x,x') \gamma^\mu.
\] 

Knowing the form of the retarded/advanced propagator for the Dirac operator, we may now tackle the singular structure of the two-point function, i.e., the Hadamard parametrix. The crucial point is that there are distributions $T^\pm_j$ fulfilling
\begin{equation}
\label{eq:T_R}
 T^+_j - T^-_j = 2 \pi i (R^\adv_j - R^\ret_j)
\end{equation}
for $j = 2 + 2k$, $k \in \N_0$, and which are of positive/negative frequency in their first argument (again, this can be made precise using the concept of the wave front set). Explicitly, in four space-time dimensions, they are given by
\begin{align}
\label{eq:T_2_4d}
 T^{\pm}_2(x,x') & = \frac{1}{2 \pi} \frac{-1}{(x-x')^2_{\pm \eps}}, \\
\label{eq:T_k_4d}
 T^{\pm}_{2 + 2 k}(x,x') & = \frac{-1}{2^{2k+1} \pi k! (k-1)!} (x-x')^{2 (k-1)} \log \frac{-(x-x')^2_{\pm \eps}}{\Lambda^2}, 
\end{align}
where $\Lambda$ is a length scale and we used the notation \eqref{eq:i_epsilon}. Note that \eqref{eq:T_R} follows straightforwardly from
\begin{align*}
 \frac{1}{x\pm i \eps} & = P \frac{1}{x} \mp i \pi \delta(x), &
 \log(x\pm i \eps) & = \log \betrag{x} \pm i \pi \theta(-x).
\end{align*}
Hence, we may now define the parametrix $h$ for the operator $P$ as
\[
 h^\pm(x,x') = \frac{1}{2\pi} \sum_{k=0}^\infty V_k(x,x') T^\pm_{2+2k}(x,x').
\]
Analogously to the construction \eqref{eq:S_ret} for the retarded/advanced propagator for the Dirac operator, we may define the parametrix $H$ for the Dirac operator as
\begin{equation}
\label{eq:H_pm}
 H^\pm(x,x') = \tfrac{1}{2} \left( - i \nabslash + i {\nabslash^*}' - 2 m \right) h^\pm(x,x')
\end{equation}
Here we took the mean of the action of the auxiliary Dirac operator \mbox{$-i \nabslash - m$} on the left and the right in order to fulfill \eqref{eq:Conj}.\footnote{We recall that the action of the auxiliary Dirac operator on the retarded/advanced propagator $\Delta^{\ret / \adv}$ gives $S^{\ret / \adv}$, regardless of whether one acts from left or right, \cf \eqref{eq:S_symmetry}.} As a straightforward consequence of \eqref{eq:T_R}, we then have
\begin{equation}
\label{eq:Hp-Hm}
 H^+(x,x') - H^-(x,x') = i S(x,x').
\end{equation}

By construction, $H^+$ is now a solution modulo smooth terms to the Dirac operator $i \nabslash - m$ in the first variable, and to its transpose in the second variable, fulfilling \eqref{eq:Comm} and the positive frequency condition. By the uniqueness result mentioned before \eqref{eq:InvariantTensors}, it follows that for any state $\omega$ with Hadamard two-point function, we have
\begin{align}
\label{eq:SmoothRemainder1}
 \omega(\psi^B(x) \bar \psi_A(y)) & = H^+(x, y)^B_A + R_A^B(x,y), \\
\label{eq:SmoothRemainder2}
 \omega(\bar \psi_A(y) \psi^B(x)) & = - H^-(x, y)^B_A - R_A^B(x,y),
\end{align}
where $R$ is smooth. That the two smooth terms on the \rhs of these equations add up to zero is a consequence of \eqref{eq:Comm} and \eqref{eq:Hp-Hm}.

We are now finally in a position to state the functions $W_i$, $1 \leq i \leq 3$ explicitly:
\begin{align*}
 W_1(x,x') & = - \tfrac{i}{2 \pi^2} \gamma_\mu (x-x')^\mu V_0(x, x'), \\
 W_2(x,x') & = \tfrac{1}{8 \pi^2} \left( i \nabslash - i {\nabslash^*}' + 2 m \right) V_0(x,x') + \tfrac{i}{16 \pi^2} (x-x')^\mu \{ \gamma_\mu, V_1(x, x') \}, \\
 W_3(x,x') & = \tfrac{1}{32 \pi^2} \left( i \nabslash - i {\nabslash^*}' + 2 m \right) V_1(x,x') + \order(x-x').
\end{align*}
From these, we obtain the relevant coinciding point limits of their derivatives:
\begin{align*}
 [W_1] & = 0, \\
 [\nabla_\mu W_1] & = - \tfrac{i}{2 \pi^2} \gamma_\mu, \\
 [\nabla_{(\mu} \nabla_{\nu)} W_1] & = 0, \\
 [\nabla_{(\mu} \nabla_\nu \nabla_{\lambda)} W_1] & = 0, \\
 [\nabla_{(\mu} \nabla_\nu \nabla_\lambda \nabla_{\rho)} W_1] & = 0, \\
 [W_2] & = - \tfrac{1}{4 \pi^2} m, \\
 [\nabla_\mu W_2] & = \tfrac{i}{16 \pi^2} \left( - \tfrac{i}{2} e \{ \gamma_\mu, \gamma^\lambda \gamma^\rho \} F_{\lambda \rho} - 2 \gamma_\mu m^2 \right), \\
 [\nabla_{(\mu} \nabla_{\nu)} W_2] & = - \tfrac{e}{24 \pi^2} \gamma^\lambda \del_{(\mu} F_{\nu) \lambda} + \tfrac{e}{24 \pi^2} \gamma_{(\mu} \del^\lambda F_{\nu) \lambda} \\
 & \hphantom{=} + \tfrac{e}{32 \pi^2} \{ \gamma_{(\mu}, \gamma^\lambda \gamma^\rho \} \del_{\nu)} F_{\lambda \rho}, \\
 [W_3] & = - \tfrac{1}{16 \pi^2} m^3 - \tfrac{i}{32 \pi^2} e m \gamma^\lambda \gamma^\rho F_{\lambda \rho} - \tfrac{1}{48 \pi^2} e \gamma^\lambda \del^\rho F_{\lambda \rho}.
\end{align*}
Note that for the last equation we used Synge's rule, i.e.,
\[
 [\nabla'_\mu V_1] = \nabla_\mu [V_1] - [\nabla_\mu V_1].
\]
These coincide with the results of Heisenberg \cite{Heisenberg34} (note, however, that there the terms with more than one $\gamma$ matrix are omitted, as they do not contribute to the current). 
When one is interested in the current $j_\xi$, these have to be traced with $\gamma_\xi$. The non-zero traces are
\begin{align}
\label{eq:TracedLimit1}
 \tr \gamma_\xi [\nabla_\mu W_1] & = - \tfrac{2 i}{\pi^2} g_{\xi \mu}, \\
\label{eq:TracedLimit2}
 \tr \gamma_\xi [\nabla_\mu W_2] & = - \tfrac{i}{2 \pi^2} g_{\xi \mu} m^2, \\
\label{eq:TracedLimit3}
 \tr \gamma_\xi [\nabla_{(\mu} \nabla_{\nu)} W_2] & = - \tfrac{e}{6 \pi^2} \del_{(\mu} F_{\nu) \xi} + \tfrac{e}{6 \pi^2} g_{\xi (\mu} \del^\lambda F_{\nu) \lambda}, \\
\label{eq:TracedLimit4}
 \tr \gamma_\xi [W_3] & = - \tfrac{e}{12 \pi^2} \del^\lambda F_{\xi \lambda}.
\end{align} 

\subsection{Comparison with Schwinger's prescription}

With our explicit knowledge of the form of the two-point function, we may now compare Dirac's method to Schwinger's \cite{Schwinger51} prescription. The latter consists in computing
\begin{multline}
\label{eq:SchwingersPrescription}
 j^\mu(x) = \tfrac{e}{2} \lim_{t\to 0_+} \left( V_0(x, x+tv) \psi^B(x + t v) \bar \psi_A(x) \right. \\
 \left. - V_0(x+tv, x) \bar \psi_A(x + t v) \psi^B(x) \right) {\gamma^\mu}_B^A,
\end{multline}
where $v$ is a future-pointing time-like unit vector. Before considering this any further, let us pause to remark that the parallel transport $V_0$ was not explicitly included in Schwinger's original work \cite{Schwinger51}. Only the later work \cite{Schwinger59} states that it has to be contained. Nevertheless, the parallel transport seems usually not to be taken into account, \cf for example the textbook and review \cite{GreinerMullerRafelski, MohrPlunienSoff98}. This leads to a breakdown of gauge invariance and spurious divergences.\footnote{Let us also mention that the textbook \cite{GreinerMullerRafelski} contains, in Section~9.7, a further mistake, stating that the above symmetric time-ordered limit should be performed as
\[
 \tfrac{1}{2} \lim_{\s{s} \to 0} \lim_{t\to +0} \left[ \bar \psi(x^0 + t, \s{x} + \s{s}) \psi(x^0, \s{x}) - \psi(x^0, \s{x}) \bar \psi(x^0 - t, \s{x} + \s{s}) \right],
\]
where $\s{s}$ is a spatial separation. But after taking the limit $t \to 0$ for fixed $\s{s} \neq 0$, the fields anti-commute, yielding
\[
 \lim_{\s{s} \to 0} \bar \psi(x^0, \s{x} + \s{s}) \psi(x^0, \s{x}),
\]
which is divergent.}

Let us now evaluate \eqref{eq:SchwingersPrescription} in a Hadamard state $\omega$. Using \eqref{eq:TracedLimit1} -- \eqref{eq:TracedLimit4}, we see that the most divergent expressions cancel, and we are left with
\begin{multline*}
 \omega(j_\mu(x)) = e [R_A^B](x) {\gamma_\mu}^A_B + \tfrac{e^2}{12 \pi^2} \del^\nu F_{\mu \nu} \lim_{t \to 0} \log t \\
+ \tfrac{e^2}{12 \pi^2} v^\lambda v^\rho \del_\lambda F_{\rho \mu} - \tfrac{e^2}{12 \pi^2} v_\mu v^\lambda \del^\rho F_{\nu \rho},
\end{multline*}
where $R$ was the smooth remainder in \eqref{eq:SmoothRemainder1}, \eqref{eq:SmoothRemainder2}. The first term on the \rhs is the one that Dirac's method gives. The second term is a logarithmic divergence, which is not canceled, unless the external current vanishes. Then also the last term vanishes. Even then, the third term remains. It shows that the definition depends on the direction in which the limit is performed.\footnote{For a coinciding point limit from a lightlike direction, this term would actually diverge. Analogous direction dependent terms were also encountered  in \cite{Christensen76} in the renormalization of the stress-energy tensor.} Its presence also spoils the conservation of the current. To eliminate this term, one would have to consider a different limit for each of the components of $j^\mu$, by choosing $v$ to be the unit vector in $\mu$ direction for the computation of the $\mu$ component (the cancellation does not depend on the fact that $v$ was originally constrained to be time-like).

The cancellation of the most severe divergences rests on the fact that these are anti-symmetric. However, this is only the case if the parallel transport is taken into account, as in \eqref{eq:SchwingersPrescription}. Otherwise, spurious divergences remain, e.g., for the leading divergence,
\begin{multline}
\label{eq:OmittedParallelTransport}
 \frac{v^\mu \gamma_\mu V_0(x+tv,x)}{t^3} -\frac{v^\mu \gamma_\mu V_0(x, x+tv)}{t^3} \\ = v^\mu \gamma_\mu \left( \frac{i v^\nu e A_\nu}{t^2} - \frac{i(v^\nu e A_\nu)^3}{6} + \order(t) \right),
\end{multline}
where for simplicity we assumed that $v^\nu A_\nu$ is constant in $v$-direction. There is also a finite remainder, which is obviously not gauge invariant. The occurrence of such spurious divergences seems to be the reason why for example in \cite{WichmannKroll56} more than one renormalization condition for the current is imposed (even though we know that there is only one ambiguity).

Finally, let us investigate the mode sum formula \eqref{eq:ModeSum} for the vacuum polarization. Assume a static external electric field, in static gauge $A_i = 0$. Let $\phi_{\s{k}}$ be the normalized modes of the Hamiltonian
\[
 H = - i \gamma^0 \gamma^i \del_i + m + \gamma^0 e A_0
\]
and $\psi_{\s{k}} = \gamma^0 \phi_{\s{k}}$. Here $\s{k}$ is a generalized momentum labeling the eigenstates (spin degrees of freedom are also included). Then the vacuum two-point function is given by
\begin{align}
\label{eq:2pt_vacuum_1}
 \omega(\psi^B(x) \bar \psi_A(y)) & = \int_{E_{\s{k}} > 0} \psi^B_{\s{k}}(\s{x}) \bar \psi_{A, \s{k}}(\s{y}) e^{- i (x^0 - y^0) E_{\s{k}}} \ud^3 \s{k} ,\\
\label{eq:2pt_vacuum_2}
 \omega(\bar \psi_A(y) \psi^B(x)) & = \int_{E_{\s{k}} < 0} \psi^B_{\s{k}}(\s{x}) \bar \psi_{A, \s{k}}(\s{y}) e^{- i (x^0 - y^0) E_{\s{k}}} \ud^3 \s{k},
\end{align}
where $x = (x^0, \s{x})$.
Choosing $v=e^0$, the unit vector in time direction and computing the vacuum polarization according to \eqref{eq:SchwingersPrescription}, we obtain
\begin{multline*}
 \omega(\rho(x)) = \frac{e}{2} \lim_{t \to 0_+} \left( \int_{E_{\s{k}} > 0} \betrag{ \psi_{\s{k}}(\s{x}) }^2 e^{-i t (E_{\s{k}} - e A_0(\s{x}))} \ud^3 \s{k} \right. \\
\left. - \int_{E_{\s{k}} < 0} \betrag{ \psi_{\s{k}}(\s{x}) }^2 e^{i t (E_{\s{k}} - e A_0(\s{x}))} \ud^3 \s{k} \right).
\end{multline*}
If it were allowed to interchange the integral and the limit here, one would obtain the mode sum formula \eqref{eq:ModeSum}. However, the integrals are not absolutely convergent, so this is not possible. In some situations (we will encounter one below), the positive/negative energy modes for a given generalized momentum $\s{k}$ have exactly opposite energy and the same modulus $\betrag{\psi_{\s{k}}}$. The mode sum formula \eqref{eq:ModeSum} would then yield a vanishing vacuum polarization. The above expression, however, gives
\[
  i e \lim_{t \to 0_+} \sin (t e A_0(\s{x})) \int_{E_{\s{k}} < 0} \betrag{ \psi_{\s{k}}(\s{x}) }^2 e^{i t E_{\s{k}}} \ud^3 \s{k},
\]
which does in general not vanish, as the integral diverges as $t \to 0$. We will discuss an explicit two-dimensional example below. This again shows that neglecting the parallel transport in Schwinger's prescription in general leads to wrong results.

\section{An explicit example in 1+1 dimensions}
\label{sec:2d}

We consider the massless Dirac field in $1+1$ dimensions, confined to the spatial interval $[-L/2, L/2]$ in the presence of a constant electric field $E$. We choose static gauge, $A_0 = E x^1, A_1 = 0$, and the $\gamma$ matrices
\begin{align*}
 \gamma^0 & = \begin{pmatrix} 0 & 1 \\ 1 & 0  \end{pmatrix}, & \gamma^1 & = \begin{pmatrix} 0 & 1 \\ -1 & 0  \end{pmatrix}.
\end{align*}
The Dirac equation
\[
 i \nabslash \psi = i \gamma^\mu (\del_\mu + i e A_\mu) \psi = 0
\]
may be written as
\begin{equation}
\label{eq:Schroedinger}
 i \del_0 \phi = i \begin{pmatrix} - 1 & 0 \\ 0 & 1 \end{pmatrix} \del_1 \phi + e E x^1 \phi,
\end{equation}
with $\phi = \gamma^0 \psi$.

In order to construct the two-point function, we proceed by the mode decomposition. The \rhs of \eqref{eq:Schroedinger} is written as $H \phi$ in the following. Eigenvectors of $H$ for eigenvalue $k$ are
\begin{align*}
 \phi_k & = \begin{pmatrix} 1 \\ 0 \end{pmatrix} e^{-\frac{i}{2} e E (x^1)^2 + i k x^1}, &
 \tilde \phi_k & = \begin{pmatrix} 0 \\ 1 \end{pmatrix} e^{\frac{i}{2} e E (x^1)^2 - i k x^1}.
\end{align*}
It follows that solutions to $i \nabslash \psi = 0$ are given by
\begin{align*}
 \psi_k & = \begin{pmatrix} 0 \\ 1 \end{pmatrix} e^{-\frac{i}{2} e E (x^1)^2 + i k (x^1 + x^0)}, &
 \tilde \psi_k & = \begin{pmatrix} 1 \\ 0 \end{pmatrix} e^{\frac{i}{2} e E (x^1)^2 + i k (-x^1 + x^0)}.
\end{align*}
To ensure self-adjointness, we have to impose boundary conditions. It does not seem to be physically reasonable to impose (anti-) periodic boundary conditions, as a particle would be accelerated around the loop, taking up energy each time. It seems more appropriate to impose bag boundary conditions \cite{MIT_bag} that ensure the vanishing of the current at the boundary:\footnote{Note the change of the sign at the two boundaries. It corresponds to contracting $\gamma$ with the inward normal.}
\begin{equation*}
 i \gamma^1 \psi|_{\pm L/2} = \pm \psi|_{\pm L/2}.
\end{equation*}
Translated into a condition on $\phi$, this ensures that $H$ is self-adjoint. The corresponding solutions are
\begin{equation*}
 \phi_n = \frac{1}{\sqrt{2L}}
\begin{pmatrix} 
 - (-1)^n \exp(\frac{i}{2} e E (\frac{L^2}{4} - x^2) + i k_n x) \\
 \exp(-\frac{i}{2} e E (\frac{L^2}{4} - x^2) - i k_n x)
\end{pmatrix} 
\end{equation*}
where $k_n = (n+\frac{1}{2}) \pi/L$ for $n \in \Z$. Before using them to construct a two-point function, let us pause to consider these solutions. Obviously, we have $\betrag{\phi_n}^2(x) = \frac{1}{2L}$, so the charge density of these modes is constant and does not depend on $n$. Hence, applying the mode sum formula \eqref{eq:ModeSum}, and always matching a positive and a negative mode, one obtains a vanishing vacuum polarization (filling positive modes or leaving holes in the negative modes would lead to a constant charge density). This seems to be the reason why in \cite{AmbjornWolfram83}, where the mode sum formula is employed, it is claimed that the vacuum polarization vanishes for massless fermions in $1+1$ spacetime dimension. As we will see, this is not correct.

Setting $\psi_n = \gamma^0 \phi_n$, we may define the two-point function as in \eqref{eq:2pt_vacuum_1}, \eqref{eq:2pt_vacuum_2}, where the integrals are replaced by a sum over $n$, and the conditions on the integration region by $k_n \gtrless 0$.
To evaluate the current or the stress-energy tensor, we have to trace over the indices $A, B$ with a $\gamma$ matrix. As these are off-diagonal, it suffices to compute the off-diagonal terms. We compute
\begin{align*}
 \omega(\psi^1(x) \bar \psi_2(y)) & = \tfrac{1}{2L} f(\tfrac{\pi}{L} (x^0-y^0+x^1-y^1-i\eps)) e^{+\frac{i}{2} e E ( (x^1)^2 - (y^1)^2)}, \\
 \omega(\psi^2(x) \bar \psi_1(y)) & = \tfrac{1}{2L} f(\tfrac{\pi}{L} (x^0-y^0-x^1+y^1-i\eps)) e^{-\frac{i}{2} e E ( (x^1)^2 - (y^1)^2)}, \\
 \omega(\bar \psi_2(y) \psi^1(x)) & = \tfrac{1}{2L} f(\tfrac{\pi}{L} (-x^0+y^0-x^1+y^1-i\eps)) e^{+\frac{i}{2} e E ( (x^1)^2 - (y^1)^2)}, \\
 \omega(\bar \psi_1(y) \psi^2(x)) & = \tfrac{1}{2L} f(\tfrac{\pi}{L} (-x^0+y^0+x^1-y^1-i\eps)) e^{-\frac{i}{2} e E ( (x^1)^2 - (y^1)^2)},
\end{align*}
where
\begin{equation*}
 f(z) = \frac{-i}{2 \sin z/2} = \frac{-i}{z} - \frac{i z}{24} + \order(z^3).
\end{equation*}
Hence,
\begin{align}
 \omega(\psi^1(x) \bar \psi_2(y)) & = \frac{-i}{2\pi} \frac{1+\frac{ieE}{2} (x^1-y^1)(x^1+y^1) - \frac{e^2 E^2}{8} (x^1-y^1)^2 (x^1+y^1)^2}{x^0-y^0+x^1-y^1-i\eps} \nonumber \\
\label{eq:omega12}
 & \hphantom{=} - \frac{i \pi}{48 L^2} (x^0-y^0+x^1-y^1) + \order((x-y)^2), \\
 \omega(\psi^2(x) \bar \psi_1(y)) & = \frac{-i}{2\pi} \frac{1-\frac{ieE}{2} (x^1-y^1)(x^1+y^1) - \frac{e^2 E^2}{8} (x^1-y^1)^2 (x^1+y^1)^2}{x^0-y^0-x^1+y^1-i\eps} \nonumber \\
\label{eq:omega21}
 & \hphantom{=} - \frac{i \pi}{48 L^2} (x^0-y^0-x^1+y^1) + \order((x-y)^2),
\end{align}
and analogously for the two-point functions $\omega(\bar \psi_A(y) \psi^B(x))$. It is then clear that Schwinger's prescription, with a coinciding point limit from the time direction, yields
\begin{equation}
\label{eq:rho_2d}
 \rho(x) = \tfrac{1}{\pi} e^2 E x^1,
\end{equation}
which has the natural interpretation of negative charges accumulating where the potential is positive, and vice versa (note that the electrostatic potential is $-A_0$). However, omitting the parallel transport, one gets a vanishing vacuum polarization $\rho(x)$. This again exemplifies the importance of including the parallel transport in \eqref{eq:SchwingersPrescription}.

The supplementary electromagnetic field produced by \eqref{eq:rho_2d} via back-reaction is
\begin{equation}
\label{eq:E_br}
 E_{\mathrm{br}} = \tfrac{1}{2 \pi} e^2 E \left[ (x^1)^2 - (L/2)^2 \right].
\end{equation}
For the model to be physically viable, this should be small compared to the external field $E$, which leads to the requirement $e L \ll 1$.

In order to apply Dirac's method, let us now compute the parametrix. In $1+1$ spacetime dimension, the distributions $T_{2k}$ are, instead of \eqref{eq:T_2_4d}, \eqref{eq:T_k_4d} given by
\[
 T^{\pm}_{2 k}(x,x') = \frac{-1}{2^{2k-1} \pi {(k-1)!}^2} (x-x')^{2 (k-1)} \log \frac{-(x-x')^2_{\pm \eps}}{\Lambda^2}
\]
for $k \geq 1$. Recalling \eqref{eq:H_pm} and \eqref{eq:V0_0} -- \eqref{eq:V1_0}, we thus see that the parametrix is given by
\begin{equation*}
 H^\pm(x,y) = \frac{-i}{2 \pi} \frac{V_0(x,y) \gamma_\mu (x-y)^\mu}{(x-y)^2_{\pm \eps}} 
\end{equation*}
up to terms that vanish as $(x-y)^2 \log (x-y)^2$ for a coinciding point limit from time-like or space-like direction. For the off-diagonal components, this means
\begin{align*}
 H^\pm(x, y)^1_2 & = \frac{-i}{2 \pi} \frac{1 - \tfrac{i}{2} e E (x^1+y^1) (x^0 - y^0) - \tfrac{1}{8} (eE)^2 (x^1+y^1)^2 (x^0 - y^0)^2}{x^0 - y^0 + x^1 - y^1 \mp i \eps}, \\
 H^\pm(x, y)^2_1 & = \frac{-i}{2 \pi} \frac{1 - \tfrac{i}{2} e E (x^1+y^1) (x^0 - y^0) - \tfrac{1}{8} (eE)^2 (x^1+y^1)^2 (x^0 - y^0)^2}{x^0 - y^0 - x^1 + y^1 \mp i \eps},
\end{align*}
again up to terms that vanish stronger than linearly in the coinciding point limit. Subtracting these from the two-point functions \eqref{eq:omega12}, \eqref{eq:omega21}, we obtain the remainder terms, \cf \eqref{eq:SmoothRemainder1}, \eqref{eq:SmoothRemainder2},
\begin{align*}
 R(x,y)^1_2 & = \tfrac{1}{4\pi} e E (x^1+y^1) - \tfrac{i}{16 \pi} e^2 E^2 (x^1+y^1)^2 (x^0-y^0-x^1+y^1) \\
 & \hphantom{=} - \tfrac{i \pi}{48 L^2} (x^0-y^0+x^1-y^1), \\
 R(x,y)^2_1 & = \tfrac{1}{4\pi} e E (x^1+y^1) - \tfrac{i}{16 \pi} e^2 E^2 (x^1+y^1)^2 (x^0-y^0+x^1-y^1) \\
 & \hphantom{=} - \tfrac{i \pi}{48 L^2} (x^0-y^0-x^1+y^1),
\end{align*}
again up to terms of higher order. Taking the coinciding point limit and contracting with $\gamma^\mu$, one finds indeed a vanishing current $j^1$ and the vacuum polarization \eqref{eq:rho_2d}. Note that filling supplementary positive energy modes (or dropping negative energy modes) just adds a constant charge density, which is independent of the energy level. The state that we have chosen is symmetric in the sense that the total charge density vanishes. In particular, this shows that the above results are not changed in a thermal state.

Contrary to Schwinger's prescription, Dirac's method also allows for a computation of the stress-energy tensor. Its classical expression is given by
\[
 T_{\mu \nu} = \tfrac{i}{2} \left( \bar \psi \gamma_{(\nu} \nabla_{\mu)} \psi - \nabla_{(\mu} \bar \psi \gamma_{\nu)} \psi \right) - \tfrac{i}{2} g_{\mu \nu} \left( \bar \psi \gamma^\lambda \nabla_\lambda \psi - \nabla_\lambda \bar \psi \gamma^\lambda \psi \right).
\]
In particular, 
\begin{align*}
 T_{00} & = \tfrac{i}{2} \left( \bar \psi \gamma_1 \nabla_1 \psi - \nabla_1 \bar \psi \gamma_1 \psi \right), \\
 T_{11} & = \tfrac{i}{2} \left( \bar \psi \gamma_0 \nabla_0 \psi - \nabla_0 \bar \psi \gamma_0 \psi \right), \\
 T_{01} & = \tfrac{i}{4} \left( \bar \psi \gamma_1 \nabla_0 \psi + \bar \psi \gamma_0 \nabla_1 \psi - \nabla_0 \bar \psi \gamma_1 \psi - \nabla_1 \bar \psi \gamma_0 \psi \right).
\end{align*}
Evaluation in our state yields
\begin{equation}
\label{eq:T_munu}
 T_{\mu \nu}(x) = - \left( \frac{\pi}{24 L^2} - \frac{e^2 E^2 (x^1)^2}{2 \pi} \right)  \begin{pmatrix} 1 & 0 \\ 0 & 1 \end{pmatrix}
\end{equation}
There are two contributions, one from the Casimir energy, and one from the interaction with the potential. The former coincides with the energy derived in \cite{SundbergJaffe}. The energy corresponding to the second term is $-\frac{1}{2}$ times the electrostatic energy density of the charge density $\rho$ in the potential $A_0$. Note that the integral over this energy density does not vanish.\footnote{There is an ambiguity in the definition of the stress-energy tensor which amounts to adding multiples of the stress-energy tensor of the external fields, which is independent of $x^1$. However, this ambiguity has to be fixed once and for all, so that it is not possible to achieve, for example, a vanishing total energy for all length $L$.} This shows that the popular mode sum expression, c.f., for example, the textbook \cite{GreinerMullerRafelski},
\[
 E_{\mathrm{tot}} = \tfrac{1}{2} \left( \sum_{-} E_n - \sum_{+} E_n \right),
\]
is not correct, as this would yield $E_{\mathrm{tot}}=0$. What may be more relevant than the energy density is the pressure at the boundaries, for which the electromagnetic contribution is positive. This may be interpreted as a weakening of the attractive force of the two boundaries due to the shielding by vacuum polarization. Now there is the puzzling fact that we do not have
\begin{equation}
\label{eq:E_P}
 - \frac{\ud E_{\mathrm{tot}}}{\ud L} = P|_{\pm L/2}.
\end{equation}
Instead, from \eqref{eq:T_munu} one obtains
\begin{align*}
 - \frac{\ud E_{\mathrm{tot}}}{\ud L} & = - \frac{\pi}{24 L^2} - \frac{e^2 E^2}{2 \pi} (L/2)^2, \\
 P|_{\pm L/2} & = - \frac{\pi}{24 L^2} + \frac{e^2 E^2}{2 \pi} (L/2)^2.
\end{align*}
The resolution of the puzzle is that one has to take into account the correction to the electromagnetic stress-energy tensor due to the modification $E \to E + E_{\mathrm{br}}$, \cf \eqref{eq:E_br}. At $\order(e^2)$, this leads to an additional term
\[
 T^{\mathrm{br}}_{\mu \nu} = \frac{e^2 E^2}{2 \pi} \left[ (x^1)^2 - (L/2)^2 \right] g_{\mu \nu}
\]
in the stress-energy tensor. Its contribution to the pressure at the boundary vanishes, but its contribution to the total energy is
\[
 E_{\mathrm{tot}}^{\mathrm{br}} = - \frac{e^2 E^2}{2 \pi} \frac{4}{3} (L/2)^3.
\]
With the inclusion of this term, the relation \eqref{eq:E_P} is indeed fulfilled.

\section{Static potentials in 3+1 dimensions}
\label{sec:4d}

In the previous section, we discussed an example where the two-point function can be computed explicitly. In general, this will not be the case. Hence, some suitable approximation scheme is needed. This is greatly simplified if the external field possesses some symmetries. In the following, we consider the case of a static potential, i.e., $A_i = 0$ and $A_0$ is time-independent.

For such a static potential, a ground state can be defined as in \eqref{eq:2pt_vacuum_1}, \eqref{eq:2pt_vacuum_2}. The practical problem is that in a numerical computation of the corresponding vacuum polarization, the finite remainder of the coinciding point limit is difficult to extract. The idea will be to write the parametrix in a form similar to \eqref{eq:2pt_vacuum_1}, so that one may integrate over the difference of the integrands. This gives an integral which is much better behaved.

Due to the static potential, we are mainly interested in the vacuum expectation value of the charge density.\footnote{It seems plausible that there are no rotational currents in the ground state, but we are not aware of a proof.} Hence, let us consider the two-point function contracted with $\gamma^0$ for a separation in time direction:
\begin{align}
 \omega(\psi^B(x+t e^0) \bar \psi_A(x)) {\gamma^0}_B^A & = \int_0^\infty \int_{E_{\s{k}}=E} \psi^B_{\s{k}}(\s{x}) \bar \psi_{\s{k}, A}(\s{x}) {\gamma^0}_B^A e^{-i t E} \ud \s{k} \ud E \nonumber \\
\label{eq:2pt_4d}
 & = \int_0^\infty f(E, \s{x}) e^{- i E t} \ud E.
\end{align}
To arrive at the second line, we performed the integration and summation over momenta and spins for energy $E$ (in the case of a radially symmetric potential, this will be a summation over angular momenta and spins). Now consider the parametrix contracted with $\gamma^0$, as one performs the coinciding point limit from the time direction. Using the results of Section~\ref{sec:ExplicitForm}, we obtain
\begin{multline*}
 H^+(x + t e^0,x)_A^B {\gamma^0}_B^A = \frac{- 2 i}{\pi^2} \frac{V_0(x+t e_0, x)}{t^3 - i \eps t^2} - \frac{i}{2 \pi^2} m^2 \frac{V_0(x+t e^0, x)}{t - i \eps} \\
- \frac{e}{12 \pi^2} \del^\lambda F_{0 \lambda} \left( \log \frac{- t^2 + i \eps t}{\Lambda^2} - 1 \right) + \order(t).
\end{multline*}
Up to the parallel transport, the first two terms also appear in the coinciding point limit of the contraction of the free two-point function, i.e., in the absence of external potentials. Denoting the sum of the first two terms by $H_0(x, t)$, we thus have
\begin{align*}
 H_0(x, t) & = e^{- i t e A_0(\s{x})} \int_0^\infty f_0(E, \s{x}) e^{- i E t} \ud E \\
 & = \int_{e A_0(\s{x})}^\infty f_0(E - e A_0(\s{x}), \s{x}) e^{- i E t} \ud E,
\end{align*}
where $f_0$ is the analog of $f$ defined in \eqref{eq:2pt_4d} in the absence of external potentials.

From the above considerations, it follows that we may write the expectation value of the charge density as
\begin{align}
\label{eq:rho_4d}
 \rho(x) & = e\lim_{t \to +0} \left( \int \left[ \chi_{[0,\infty)}(E) f(E, \s{x}) \right. \right. \\
 & \qquad \qquad \qquad \left. - \chi_{[e A_0(\s{x}), \infty)}(E) f_0(E- e A_0(\s{x}), \s{x}) \right] e^{ - i E t} \ud E \nonumber \\
 & \qquad \qquad \left. + \frac{e}{12 \pi^2} \del^\lambda F_{0 \lambda} \left( \log \frac{- t^2 + i \eps}{\Lambda^2} - 1 \right) \right). \nonumber
\end{align}
The second term in the integral has the natural interpretation of subtracting the vacuum contribution at the adjusted local energy level. Note that if the state $\omega$ should be defined by a different choice of zero-point energy, one simply has to adjust the characteristic function multiplying $f(E, \s{x})$ accordingly. Also note that in some cases it may be advantageous to work with the two-point function for $\bar \psi \psi$ instead of the one for $\psi \bar \psi$. Then the sign in \eqref{eq:rho_4d} is changed, and the domains of integration are from $-\infty$ to $0$ (respectively $e A_0(\s{x})$). Finally, we note that in the absence of an external charge density, when the logarithmic divergence is absent, it may indeed be possible to perform the limit $t \to 0$ in the integrand and have a numerically stable integral. However, this integral will in general still not be absolutely convergent, but oscillatory.

We exemplify the above considerations with the Coulomb potential. To linear order in the external potential, the vacuum polarization was calculated, using Dirac's method, by Uehling \cite{Uehling35}. Higher order corrections were later computed, using Schwinger's prescription, by Wichmann and Kroll \cite{WichmannKroll56}, \cf also the review \cite{MohrPlunienSoff98} and references therein. However, their calculation did not take the parallel transport in \eqref{eq:SchwingersPrescription} into account, which led to a divergent term at first order in the external potential, corresponding to the first term on the \rhs of \eqref{eq:OmittedParallelTransport}. Being of the same order in the external field, this term was then identified with the Uehling term. Also at third order in the external potential some regularization is needed in the calculation of Wichmann and Kroll. Their procedure seems to boil down to also subtract the second, finite, term on the \rhs of \eqref{eq:OmittedParallelTransport}. So in the end, gauge invariance is restored, albeit in a rather intransparent way.

\begin{figure}
\includegraphics{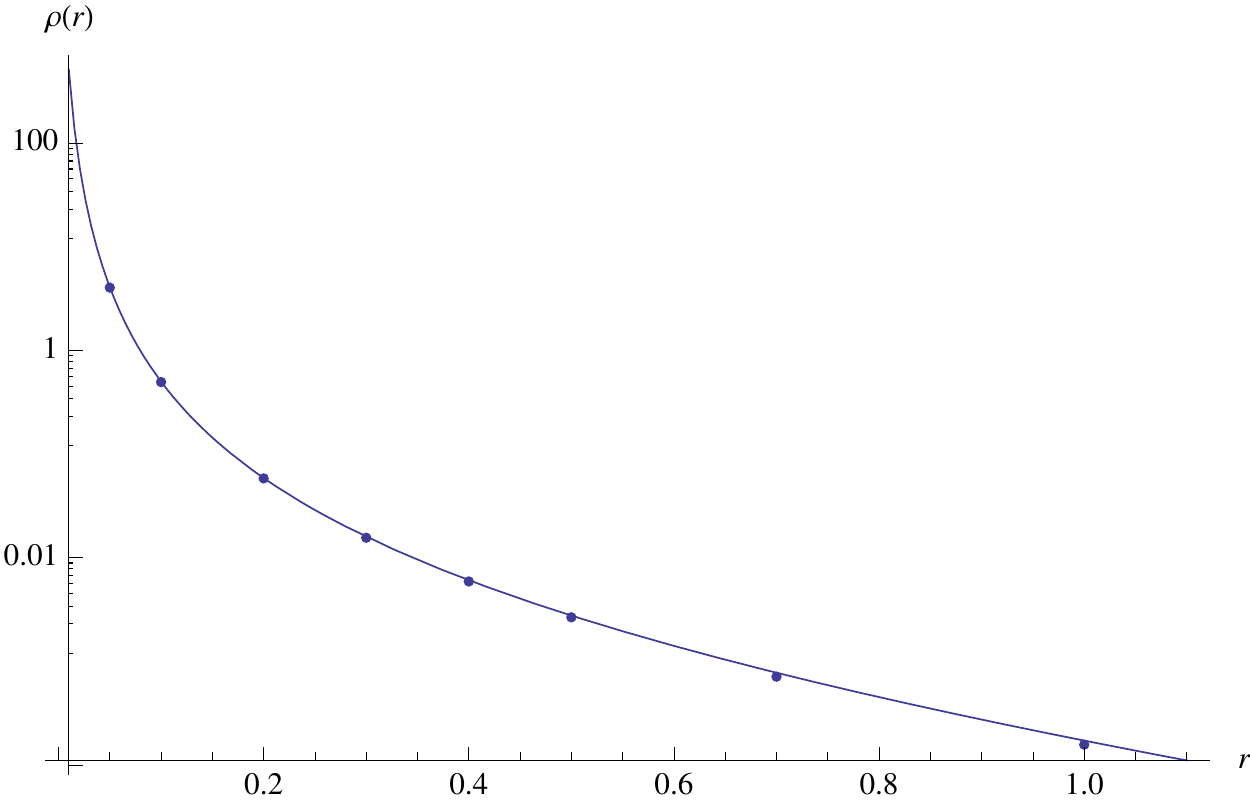}
\caption{The vacuum polarization in a Coulomb potential (dots), computed using \eqref{eq:rho_4d}, as a function of the radius (in units of the electron Compton wavelength) for $\alpha Z = 0.1$. As a comparison, the vacuum polarization according to Uehling is also shown.}
\label{fig:Coulomb}
\end{figure}

In order to illustrate the usefulness of the expression \eqref{eq:rho_4d}, we use it to compute the vacuum polarization in the Coulomb potential. The explicit form of the modes of the continuous spectrum can be found for example in \cite{Rose}. As the external charge density vanishes except at the origin, the second term on the \rhs of \eqref{eq:rho_4d} is absent, and the limit $t \to 0$ may be performed inside the integral. We consider the state obtained by filling up the negative continuum, i.e., by considering a completely ionized nucleus. Figure~\ref{fig:Coulomb} shows the charge density $\rho(r)$ for $\alpha Z = 0.1$, as a function of the radius. We see a nice agreement with the result of Uehling \cite{Uehling35} for the vacuum polarization linearized in the external field. These numerical results mainly serve as an illustration that the above method can be practically applied. The concrete numerical implementation can certainly be improved considerably. In particular it should be possible to compute the higher order corrections to Uehling's result. In a separate publication \cite{SauterPotential}, a time-dependent homogeneous electric field is treated in an analogous way, with agreement to well-established results both in the perturbative and the non-perturbative regime.

\subsection*{Acknowledgements}
It is a pleasure to thank Michael Bordag, Christian Brouder, and Micha{\l} Wrochna for helpful discussions and hints to the literature. Parts of this work were done while J.Z.\ was at the Faculty for Physics of the University of Vienna, where he was supported by the Austrian Science Fund (FWF) under the contract P24713.


\end{document}